# Anion vacancies as a source of persistent photoconductivity in II-VI and chalcopyrite semiconductors

Stephan Lany and Alex Zunger

*National Renewable Energy Laboratory, Golden, CO 80401*

## ABSTRACT

Using first-principles electronic structure calculations we identify the anion vacancies in II-VI and chalcopyrite Cu-III-VI$_2$ semiconductors as a class of intrinsic defects that can exhibit metastable behavior. Specifically, we predict persistent *electron* photoconductivity (*n*-type PPC) caused by the oxygen vacancy V$_O$ in *n*-ZnO, and persistent *hole* photoconductivity (*p*-type PPC) caused by the Se vacancy V$_{Se}$ in *p*-CuInSe$_2$ and *p*-CuGaSe$_2$. We find that V$_{Se}$ in the chalcopyrite materials is amphoteric having two "negative-*U*" like transitions, i.e. a double-donor transition $\varepsilon(2+/0)$ close to the valence band and a double-acceptor transition $\varepsilon(0/2-)$ closer to the conduction band. We introduce a classification scheme that distinguishes two types of defects (e.g., donors): type-$\alpha$, which have a defect-localized-state (DLS) in the gap, and type-$\beta$, which have a resonant DLS within the host bands (e.g., conduction band). In the latter case, the introduced carriers (e.g., electrons) relax to the band edge where they can occupy a perturbed-host-state (PHS). Type $\alpha$ is non-conducting, whereas type $\beta$ is conducting. We identify the *neutral* anion vacancy as type-$\alpha$ and the doubly positively *charged* vacancy as type-$\beta$. We suggest that illumination changes the charge state of the anion vacancy and leads to a crossover between $\alpha$- and $\beta$-type behavior, resulting in metastability and PPC. In CuInSe$_2$, the metastable behavior of V$_{Se}$ is carried over to the (V$_{Se}$-V$_{Cu}$) complex, which we identify as the physical origin of PPC observed experimentally. We explain previous puzzling experimental results in ZnO and CuInSe$_2$ in the light of this model.

## I. INTRODUCTION

Impurities and defects in semiconductors create defect-localized-states (DLS) that are located either in the gap, or resonate within the continuum of the host bands. In general, different host/impurity combinations may be divided into two categories: those that create localized gap states, and those that create only resonant states. For example, a DLS located within the gap and below the conduction band minimum (CBM, "$\alpha$-type behavior") is created for most 3*d* impurities in III-V semiconductors [1], nitrogen [2, 3, 4] and oxygen [3, 5] in GaP, oxygen in ZnTe [6, 7], hydrogen in MgO [8], or oxygen vacancies in Al$_2$O$_3$ [9]. On the other hand, the DLS is located above the CBM ("$\beta$-type behavior") for nitrogen in GaAs [3, 4, 10], oxygen in ZnS or ZnSe [11], hydrogen in ZnO [12], oxygen vacancies in In$_2$O$_3$ [9], as well as all classic, hydrogenic donors in semiconductors [13]. In the "DLS-below-CBM" case, one expects a deep level that has localized wavefunctions and that consequently responds only weakly to external perturbations such as pressure or temperature. Even if such a defect brings-in carriers, it usually does not lead to conductivity due to the localized nature of the state and the ensuing high activation energy. Conversely, in the β-type situation where the DLS is resonant with the conduction band, electrons in this level will drop to the conduction band minimum (CBM), and occupy a perturbed-host-state (PHS, Ref. [4]) rather than the DLS. In this case, a delocalized, hydrogenic-like state is created which is conductive and responds more strongly to external perturbations, due to its host-band-like character.

There are unique cases where the *same* impurity in a solid can assume both type-α and type-β behavior. This is the case for the "*DX* center" [14], where a donor atom such as Te in AlGaAs [15] or Cl in CdZnTe [16] exhibits the conducting "DLS-above-PHS" β-type behavior in its normal substitutional configuration, but after a large lattice relaxation involving bond breaking [17], it exhibits the insulating "DLS-below-CBM" α-type behavior. The hallmark of the *DX* centers is the phenomenon of persistent photoconductivity (PPC), arising from the light induced configuration change from the non-conducting ground state to the metastable conductive state.

In this paper, we discuss the interesting case where *intrinsic* defects (not impurities) exhibit both α-type and β-type behavior in the same solid, depending on their charge state. Remarkably, this occurs already for one of the simplest point-defects, namely the anion vacancy in II-VI and in their ternary analogues, i.e. in the I-III-VI$_2$ chalcopyrites. Through first principles total energy calculations, we show that the transition between α- and β-type configurations results in metastable behavior of these defects. Specifically, we predict that the oxygen vacancy $V_O$ leads to persistent *electron* photoconductivity (*n*-type PPC) in *n*-ZnO observed recently after proton irradiation [18], whereas the Se vacancy $V_{Se}$ causes persistent *hole* photoconductivity (*p*-type PPC) in *p*-CuInSe$_2$ or *p*-CuGaSe$_2$, constituting the unusual case where a donor-like defect causes *p*-type PPC. Thus, our results shed new light on hitherto unknown physical origin of the phenomenon of light-induced metastability in chalcopyrite CuInSe$_2$ based photovoltaic devices exhibiting *p*-type PPC [19, 20, 21, 22, 23]. Here, this light-induced effect leads to performance *enhancement* [19, 21, 23], in stark contrast to "light-induced degradation" of other solar cells, e.g. Staebler-Wronski *degradation* in *a*-Si [24].

## II. BASIC PHYSICS OF α-TYPE AND β-TYPE BEHAVIOR

The distinction between these two types of behaviors can be appreciated from simple two-level models in which the impure system is described via the interaction between a ideal (unrelaxed) vacancy (vac) state $\varepsilon_{vac}^{\lambda,\gamma}$ in the pure host crystal and the orbitals $\varepsilon_I^{\lambda,\gamma}$ of the isolated impurity (I) atom (Fig.1) [3, 25]. Here, we regard only the dilute limit, where no interaction between impurity states exists. The vacancy of the pure host crystal can occur on either sublattice λ (i.e. λ = cation or λ = anion). Different vacancy levels can be constructed from the dangling bond hybrids of the neighboring host atoms, according to the irreducible representations γ of the corresponding λ site symmetry (e.g., γ = $a_1$ or γ = $t_2$ in the $T_d$ point group symmetry of the zincblende structure). We define $\varepsilon_{vac}^{\lambda,\gamma}$ in Fig. 1 as the energy of the λ-site vacancy state with symmetry γ. Similarly, the isolated impurity atom on site λ has symmetry-adapted orbitals with energy $\varepsilon_I^{\lambda,\gamma}$. The two unperturbed states $\varepsilon_{vac}^{\lambda,\gamma}$ and $\varepsilon_I^{\lambda,\gamma}$ are shown in Fig. 1 as dotted lines, representing varying impurities in one host system.

Vacancy and impurity states of same symmetry γ interact and repel each other, creating bonding and antibonding states (red solid lines in Fig. 1), as a function of the perturbation strength $\Delta V = \varepsilon_I^{\lambda,\gamma} - \varepsilon_H^{\lambda,\gamma}$, measured by the difference of atomic orbital energies between the impurity (I) atom and the host (H) atom it replaces [25]. As one expects, for extremal values of $\Delta V$ the energy of the impurity states saturate at the value of the host vacancy level. Indeed, when one regards the vacancy itself as the impurity, the energies $\varepsilon_{vac}^{\lambda,\gamma}$ of the vacancy states are not altered by interaction with an impurity state $\varepsilon_I^{\lambda,\gamma}$; as discussed below, the energies of the vacancy states can depend strongly on atomic relaxation, however.

In order to illustrate the range of physical behaviors that is implied with the α- vs. β–behavior, we now add schematically for the specific case λ = *anion* and γ = $a_1$ the position of the host band edges in Fig. 1 (blue horizontal solid lines), and integrate in the figure the examples GaP:O, ZnTe:O, GaAs:N, GaAs:Te, and the neutral anion vacancy $V_O^0$ in ZnO, which we discuss in detail below. In all those examples except for the vacancy, the occupied lower energy *bonding* state represents the impurity atomic-*s* state, e.g. O-2*s*, that lies deep below the valence band maximum (VBM). What determines the physical behavior of the impurity is the energetic position of the DLS created by the *antibonding* combination of the vacancy and the impurity state. We see in Fig. 1 that the α-type situation is realized for strongly attractive (i.e. electronegative) impurities with $\varepsilon_I^{\lambda,\gamma} \ll \varepsilon_H^{\lambda,\gamma}$. The host/impurity combinations of Fig. 1 produce the following different physical properties: (*i*) *α-type behavior for substitution by a donor*: For example, GaP:O creates an *occupied* (closed circle in Fig. 1) deep non-conducting donor state (*ii*) *α-type behavior for isoelectronic substitution*: ZnTe:O creates an *unoccupied* (open circle in Fig. 1) acceptor state in the upper part of the gap. (*iii*) *β-type behavior for isoelectronic substitution*: For example, GaAs:N creates a *unoccupied* resonant DLS in the conduction band, but no transition level in the band gap. (*iv*) *β-type behavior for substitution by a donor*: GaAs:Te creates an *occupied* shallow conducting donor state. Here, the donor electron occupies the perturbed-host-state (PHS) which is created below the CBM by the long-range, screened Coulomb potential introduced by the donor substitution.

In this paper, we will show that anion vacancies in II-VI and Cu-III-VI$_2$ compounds can exhibit both type α and β behavior: The neutral oxygen vacancy $V_O^0$ in ZnO shows α-type behavior, where the doubly occupied DLS is below the CBM in the band gap (Fig. 1). However, the doubly charged state $V_O^{2+}$ (not shown in Fig. 1) has an unoccupied $a_1^0$ vacancy level above the CBM (β-type behavior with a resonant DLS), and creates a PHS below the CBM. Thus, $V_O$ in ZnO can assume a non-conducting configuration, where both electrons are in the deep DLS (α-type), and a conducting configuration, where the DLS is empty and the electrons occupy the shallow PHS (β-type).

Besides its use as a general classification, the distinction between scenario α and β (Fig. 1) has a technical implication on total-energy *supercell* calculations: Due to the localized nature of an α-type defect state, e.g. $V_O^0$ in ZnO or in Al$_2$O$_3$, the effects of impurity-impurity interaction and impurity band formation in generally weak and can be rapidly suppressed by increasing the size of the supercell [26]. However, due to the delocalized nature of a β-type defect, however, a highly dispersive impurity band [4, 9] forms already at rather low concentrations, leading to a Moss-Burstein shift to higher energies if the PHS is occupied. For example, Astala and Bristowe [27], as well as Buban and co-workers [28] calculated $V_O^0$ in SrTiO$_3$ and observed strong changes in total energy and atomic relaxation and with increasing cell size (up to 320 atoms). While long-range defect image interaction and long-range relaxation effects were invoked in Ref. [27] and Ref. [28], respectively, we note that such behavior could also be explained by the Moss-Burstein like band-filling effect if $V_O^0$ in SrTiO$_3$ assumes the β-type scenario: in the limit of large supercells, the two electrons occupy the PHS, while the higher-energy DLS is unoccupied ($a_1^0$ configuration). With decreasing supercell size, the Moss-Burstein effect leads to an upward shift of the Fermi energy, so that the DLS becomes partly occupied ($a_1^x$, $0 < x < 2$). Taking into account that the occupancy of the DLS controls the large lattice relaxations (viz. Ref. [29] and discussion below), both the change in total energy and in atomic relaxation can be explained by this interpretation.

### III. METHOD OF CALCULATION

We use first-principles supercell calculations in order to determine the atomic structure, the single-particle defect levels, the defect formation energy $\Delta H$, as well as thermal (equilibrium) and optical (vertical) transition energies ε for the anion vacancy in II-VI and Cu-III-Se$_2$ compounds. Here, the formation energy of the vacancy is defined as

$$\Delta H_{D,q}(E_F, \mu) = (E_{D,q} - E_H) + (\mu_X^{elem} + \Delta\mu_X) + q(E_V + \Delta E_F), \qquad (1)$$

where $E_{D,q}$ is the total energy of the semiconductor with the defect D (here, the vacancy) in charge state $q$, and $E_H$ is the energy of the pure host. The second term describes the chemical reservoir in equilibrium, where the chemical potential $\mu_X = \mu_X^{elem} + \Delta\mu_X$ of the removed anion X is given with respect to the elemental phase. For the elemental reference $\mu_X^{elem}$, we choose the solid phase except for oxygen, where we choose the isolated O$_2$ molecule, i.e. $\mu_O^{elem} = \frac{1}{2}\mu_{O_2}$. The third term in Eq. (1) is the energy of the electron reservoir, i.e. the Fermi energy $E_F = E_V + \Delta E_F$ given with respect to the energy $E_V$ of the VBM. Note that the boundaries of the allowed range of $\mu_X$ are defined by the stability condition of the host material, e.g. $\Delta\mu_{Zn} + \Delta\mu_O = \Delta H_f(\text{ZnO})$, where $\Delta H_f$ is the calculated compound formation energy. In case of chalcopyrite, the range of chemical potentials is further restricted by formation of competing phases [30].

The thermal transition energy $\varepsilon(q/q')$ is the Fermi energy where the lowest-energy charge state changes from $q$ to $q'$ as $\Delta E_F$ rises in the gap, i.e. where $\Delta H(q, E_F) = \Delta H(q', E_F)$. From Eq. (1), it follows

$$\varepsilon(q/q') = \frac{E_D(q') - E_D(q)}{q - q'}. \qquad (2)$$

We also calculate the optical (vertical) transition energies, where the following transitions are considered: (*i*) Excitation of electrons from an occupied defect state to the CBM; the charge state $q$ of the defect changes by +1 during this transition. (*ii*) Photon emission due to the decay of an electron at the CBM into an unoccupied defect state; the charge state $q$ changes by −1. (*iii*) Excitation of an electron from the VBM into the unoccupied defect state; the charge state $q$ changes by −1. (*iv*) Photon emission due to the decay of an electron in the defect state with a hole at the VBM; the charge state $q$ changes by +1.

Keeping the VBM as reference, i.e. taking $\Delta E_F = 0$ for calculation of $\Delta H$ in Eq. (1), we calculate all optical transition energies from total energy differences. We fix the relaxed atomic positions of the initial state, assuming the Franck-Condon principle. Thus, transitions (*i*) and (*ii*) which communicate with electrons (*e*) at the CBM are calculated as (with $n = +1$ and $n = -1$, respectively)

$$\varepsilon_o(q/q+n;ne) = \Delta H(q+n) - \Delta H(q) + nE_g, \tag{3}$$

where $E_g$ is the (experimental) band gap energy. Similarly, transitions (*iii*) and (*iv*) which communicate with holes (*h*) at the VBM are calculated as (with $n = +1$ and $n = -1$, respectively)

$$\varepsilon_o(q/q-n;nh) = \Delta H(q-n) - \Delta H(q). \tag{4}$$

Note that only the ground-state energies for the respective charge states enter in Eqs. (3) and (4). Excitation (absorption, $n = +1$) energies are calculated as positive energies from Eqs. (3) and (4), while recombination (emission, $n = -1$) energies are calculated as negative. Such calculated optical transition energies have been used in [29] to explain the experimentally observed absorption energies in case of the color centers $V_S^+$ in ZnS and $V_{Se}^+$ ZnSe.

The total energies and atomic forces were calculated in the pseudopotential-momentum space formalism [31] within the local density approximation (LDA) of density functional theory. We use the Ceperley-Alder LDA exchange correlation potential as parameterized by Perdew and Zunger [32] and projector augmented wave potentials [33] as implemented in the VASP code [34]. The energy cutoff in the plane-wave expansion was up to 400 eV. Most results were obtained from supercells with 64 lattice sites for the zincblende and chalcopyrite structures, and with 72 sites for the wurzite ZnO, using experimental lattice constants. For total-energy calculation, Brillouin zone integrations were performed on a Γ-centered 3×3×3 mesh (3×3×2 in ZnO) using the improved tetrahedron method, but for relaxation a reduced k-mesh was used.

The well-known LDA band gap error is corrected by acknowledging that the cation $d$ states in II-VI and chalcopyrites are too shallow on account of their strong, spurious self-interaction [32b]. Thus, we use the LDA+U method [35] to lower the Zn-3$d$ ($U = 7$eV) and Cu-3$d$ ($U = 5$eV) bands so as to yield agreement of the $d$-like density of states with Zn and Cu photoemission data in II-VI and chalcopyrite [36, 37, 38, 39], respectively. This $d$-band lowering weakens the $p$-$d$ repulsion with the anion-$p$ orbitals [40], and lowers the energy of the VBM. Thus, $E_V$ in Eq. (1) is corrected by $\Delta E_V = -0.77, -0.34 -0.28$, and $-0.20$ eV in ZnO, ZnS, ZnSe, and ZnTe, respectively, and by $\Delta E_V = -0.37$ eV in CuGaSe$_2$ and CuInSe$_2$. Still, the LDA+U band gaps, e.g. 1.53 eV in ZnO and ~0 eV in CuInSe$_2$ are much smaller than the experimental gaps (3.37 eV in ZnO [41] and 1.04 eV in CuInSe$_2$ [42]). The remaining discrepancy is accommodated by shifting the CBM upwards (by $\Delta E_C$). In practice, we take this correction into account by using the experimental band gap energy $E_g$ in Eq. (3) and for the allowed range $0 \leq \Delta E_F \leq E_g$ of $\Delta E_F$ in Eq. (1). In case of β-type defects that introduce electrons in the PHS, i.e. in case of shallow donors, a correction $n\Delta E_C$ applies to $\Delta H$ when the PHS is occupied by $n$ electrons. This is because the PHS shifts along with the CBM during band gap correction. Similarly, a correction $-n\Delta E_V$ applies to shallow acceptors when the acceptor state (valence-band-like PHS) is occupied by $n$ holes. In case of α-type defects, where the electrons or holes occupy the DLS, the latter two corrections are not applied, because the DLS does not strictly follow the band edges like the PHS does.

Several further corrections to the LDA calculated energies were applied, as discussed in more detail in another publication [43]: First, Moss-Burstein like band filling effects (cp. section II) occur when electrons or holes occupy a perturbed-host-state forming an impurity band in the supercell calculation. Such band-filling effects have also to be taken into account when, e.g., an unoccupied defect level drops below the VBM in the course of atomic relaxation, thus releasing holes to the valence band (cp. section V). Second, the potential alignment between a charged defect calculation and the perfect host crystal is needed in order to correct for the effect of the compensating background charge. Third, the spurious interaction of periodic image charges is corrected to $O(L^{-5})$ [44], where $L$ is the linear supercell dimension. The present scheme of treating finite supercell size effects and LDA errors has shown to yield good results for the optical transition energies of the color center in ZnS and ZnSe [29], when compared with experimentally observed absorption energies.

### IV. ANION VACANCIES IN ZnO AND OTHER II-VI

Having introduced in section II the model of α- vs. β-type behavior for the general case, where the impurity atomic orbitals interact with the states of the ideal vacancy, we now turn to the levels of the relaxed anion vacancy in ZnO and other II-VI compounds. As we have shown previously [29], the energies of the vacancy orbitals (e.g., $a_1$ and $t_2$ in $T_d$ symmetry) depend strongly on atomic relaxation of the cation neighbors. This is because these orbitals are constructed themselves from combinations of the dangling bond hybrids centered at each of the four cation neighbors, the totally symmetric $a_1$ state being bonding-like. An inward relaxation increases the overlap between the dangling bond hybrids, thus lowering the energy of the $a_1$ state. Generally, the occupation of this bonding-like state in the neutral charge state $V_X^0$ ($a_1^2$ configuration) leads to atomic relaxation that brings the neighboring cations closer, while in the ionized $V_X^{2+}$ state ($a_1^0$), the cations relax outwards to a near-planar atomic configuration [29].

#### a. Oxygen vacancies cause *n*-type PPC in ZnO

Figure 2a shows the angular-momentum decomposed vacancy-site local density of states (DOS) of the relaxed neutral ($V_O^0$) and charged ($V_O^{2+}$) oxygen vacancy in ZnO, and the DOS of the host semiconductor. We see that the relaxed neutral vacancy $V_O^0$ in ZnO (Fig. 2a, middle panel) has an occupied $a_1$ symmetric (*s*-like) state in the band gap. (Note that the correction $\Delta E_V$ described in section III shifts this level to somewhat higher energies with respect to the VBM

*Stephan Lany and Alex Zunger, submitted to Phys. Rev. B, preprint date 1/14/05.*than indicated in Fig. 2a.) Thus, $V_O^0$ assumes the α-type "DLS-below-CBM" behavior, outlined in the Introduction and in Fig. 1. The energy of this state is deepened by the *inward* relaxation of the nearest-neighbor Zn atoms towards the vacancy site, leading to a local symmetry with one Zn atom at 1.78 Å from the vacancy, and three Zn atoms at 1.83 Å from the vacancy. The resulting average Zn-Zn distance is $d_{Zn-Zn} \approx 3.0$ Å, compared to $d_{Zn-Zn} \approx 3.23$ Å in the unperturbed ZnO lattice.

When forming the ionized 2+ state $V_O^{2+}$, in which the DLS is unoccupied, the Zn neighbors relax instead *outwards*, leading to a configuration with a much larger $d_{Zn-Zn} \approx 4.0$ Å. As a result of the outward relaxation, the $a_1$ state is shifted to higher energies, becoming a broad *unoccupied* resonance in the conduction band at $\sim E_C + 0.4$ eV (Fig. 2a, bottom panel). Thus, the ionized $V_O^{2+}$ creates a PHS below the DLS, constituting β-type behavior.

Figure 3a summarizes schematically the calculated energy levels discussed above. Upon photoexcitation, the electrons in the occupied DLS of $V_O^0$ (Fig. 3a, left hand side) are promoted into the conduction band, creating the ionized state $V_O^{2+}$ (Fig. 3a, right hand side). This process is described by the semi-quantitative [45] configuration coordinate diagram, shown in Fig. 3b: First, both electrons occupy the deep and non-conducting α-type state $V_O^0$. The average Zn-Zn distance, serving as the reaction coordinate, is small in this configuration ($d_{Zn-Zn} \approx 3.0$ Å). Optical excitation of the $V_O^0$ ground state to the $V_O^+ + e$ exited state occurs at the energy $\varepsilon_o(0/+;e) = +2.83$ eV. This transition creates $V_O^+$ having a *singly* occupied DLS within the band gap, i.e. being α-type. The equilibrium Zn-Zn distance of $V_O^+$ is estimated [45] as $d_{Zn-Zn} \approx 3.2$ Å (Fig. 3b). Due to the single occupancy of the DLS, the $V_O^+$ state is active in electron paramagnetic resonance (EPR), and is indeed observed in EPR experiments under illumination [46, 47]. A second excitation $V_O^+ \rightarrow V_O^{2+} + e$ occurs at $\varepsilon_o(+/2+;e) \approx +2.4$ eV, producing the $V_O^{2+}$ state with $d_{Zn-Zn} = 4.0$ Å (Fig. 3b). Following this large outward relaxation, the DLS moves upward, becoming resonant inside the conduction band (Fig. 3a, right hand side). Consequently, the photoexcited electrons occupy the lower energy PHS rather than the DLS, i.e. the $V_O^{2+}$ vacancy assumes β-type behavior (Fig. 3a, right hand side). The electrons in the PHS are now only shallowly bound through the screened Coulomb potential.

Figure 4a shows as a function of the Fermi energy $E_F$ the formation energies of the light-induced metastable configuration of $V_O^0$ and $V_O^+$ (dashed lines), relative to the corresponding formation energies in the respective equilibrium stable configuration (solid lines) [48]. The transition energies in the metastable β-type configuration (open circles in Fig. 4a) are close to the CBM [49], so that this configuration is conductive. These shallow binding energies of the electrons in the PHS are also schematically indicated in Fig. 3b by the vertical displacement of the energy curves at $d_{Zn-Zn} = 4.0$ Å. The reaction path including both excitations in Fig. 3b leads to the light-induced transition

$$V_O^0 \rightarrow V_O^{2+} + 2e, \tag{5}$$

which results in metastable configuration change from non-conducting α- to conducting β-behavior, constituting persistent *electron* photoconductivity (*n*-type PPC).

Since the atomic relaxation from the large Zn-Zn distance (β-type) to the small Zn-Zn distance (α-type) is controlled by the occupancy of the DLS [29], the light-induced β-type configuration with an empty DLS ($a_1^0$) is (meta)stable against the depopulation of electrons from the PHS into the deep ground state with an energy barrier $\Delta E$ (Fig. 3b). Unfortunately, the explicit calculation of $\Delta E$ in LDA is prevented by the problem outlined in footnote [45]. This barrier is located at $d_{\text{Zn-Zn}} \approx 3.7$ Å, at which distance the DLS drops below the (experimental) CBM. Once the DLS captures electrons from the conduction band after thermal activation across $\Delta E$, it becomes occupied, and atomic relaxation leads to the α-type deep ground state $V_O^0$ with small $d_{\text{Zn-Zn}}$, and PPC is lost.

**b. Transition and formation energies of $V_O$ in ZnO: Comparison with experiment**

We compare our calculated optical transition energies of the oxygen vacancy with recent experimental findings obtained by optically detected magnetic resonance (ODMR) and photoluminescence excitation (PLE) spectroscopy studies [50, 51]. The green emission of ZnO around 2.45 eV was identified by ODMR as arising from a $S = 1$ spin triplet state attributed to $V_O$ [50]. In addition, the PLE experiments in Ref. [51] have shown that the $S = 1$ state can be produced by sub-band-gap light of 3.1 eV. The latter energy is only slightly higher than our calculated optical transition energy $\varepsilon_o(0/+;e) = +2.83$ eV corresponding to the $V_O^0 \rightarrow V_O^+ + e$ transition. Thus, we interpret the experimental results in the following way: First, the neutral $V_O^0$ is excited to $V_O^+$, where the electron is released to the conduction band [the $\varepsilon_o(0/+;e)$ transition in Fig. 3b]. Due to its charge, $V_O^+$ creates a hydrogenic-like PHS below the CBM, which can be populated by a conduction band electron. The electron in the PHS aligns its spin with the electron in the DLS of $V_O^+$ ($a_1^1$) to form the triplet state. Finally, the recombination of the electron from the PHS into the DLS causes the green emission at 2.45 eV, corresponding to the theoretical $\varepsilon_o(+/0;-e) \approx -2.1$ eV transition in Fig. 3b [52]; note that the radiative recombination of the triplet state involves a spin flip, leading to a rather long lifetime. The observation of a $S = 1$ state in the recent experiments using moderately *n*-type samples contrasts with the observation of a $S = 1/2$ state in the classic experiments [46, 47], where the samples were compensated either by additional Li-doping [47] or by the simultaneous formation of oxygen and zinc vacancies during high energy electron irradiation [53]. This behavior is easily understood in our model where the α-type $V_O^+$ forms the triplet state by binding a conduction band electron in *n*-ZnO, but not in the compensated samples, where photoexcited electrons can recombine with acceptor states.

Whether $V_O$ is present in ZnO in relevant concentrations, depends on the thermodynamic boundary conditions during crystal growth, i.e. on the chemical potentials $\Delta\mu$, and on the position of the Fermi energy $\Delta E_F$ in the gap, which depends on doping. For the limit of maximally oxygen-poor conditions under which $\Delta H(V_O)$ is lowest, i.e. $\Delta\mu_O = \Delta H_f(\text{ZnO}) = -3.53$ eV, Fig. 4a shows as a function of $\Delta E_F$ the formation and transition energies for the different charge states in their equilibrium configuration (solid lines and closed circles) [54]. We see that $\Delta H(V_O^{2+}) = 0$ at $E_F \approx E_V + 1$ eV, which means that $V_O^{2+}$ forms spontaneously and acts as hole killer when $E_F$ approaches the VBM. The formation energy of $V_O^0$, being the equilibrium stable state in *n*-ZnO, is $\Delta H(V_O^0) = 1.14$ eV under oxygen-poor conditions, which yields an equilibrium concentration of, e.g., $c(V_O) \approx 10^{17}$ cm$^{-3}$ at $T = 1000$ K. The rather high formation energy $\Delta H(V_O^0) = 4.67$ eV under oxygen-rich growth conditions ($\Delta\mu_O = 0$) indicates that significant concentrations of $V_O$ would not occur under these conditions, except if $E_F$ would be close to the VBM (i.e., in *p*-type ZnO), in which case $\Delta H$ of $V_O$ is lowered by forming the ionized $V_O^{2+}$ state (cp. Fig. 4a).

The calculated transition energies of Fig. 4a show the "negative-*U*" behavior typical for anion vacancies in II-VI [55], i.e. the first ionization energy $\varepsilon(+/0)$ is deeper (farther from the CBM) than the second ionization energy $\varepsilon(2+/+)$. This leads to a calculated $\varepsilon(2+/0)$ transition at $E_F = E_V + 1.60$ eV from the 2+ state directly into the neutral state, as shown in Fig. 4a. Owing to this deep transition level (closed circle in Fig. 4a), the equilibrium stable α-type configuration of $V_O^0$ is not expected to provide for *n*-type doping. While hydrogen has recently been identified as a source of *n*-type conductivity in ZnO [56, 12], an additional contribution could, however, arise from PPC of $V_O$ under illumination, due to the shallow transition energies (open circles in Fig. 4a) of the β-type metastable configuration. Note that the back-transition from the conducting (β-type) to the non-conducting (α-type) configuration requires thermal activation across $\Delta E$ (Fig. 3b) and *simultaneous* capture of an electron. Thus, the lifetime of the metastable configuration and, hence, its concentration during constant illumination, increases at low temperature and low electron concentration. Our results further suggest that the recently observed [18] persistent increase of capacitance after illumination of proton-irradiated ZnO results from PPC of oxygen vacancies created by the irradiation.

Comparing our results on $V_O$ with other first-principles calculations, we note that only the energies of the ground state configurations (solid lines in Fig. 4a) have been calculated so far, and that there exists quite a spread in the values of the calculated formation and transition energies in literature. For example, $\Delta H(V_O^0)$ in the oxygen-rich regime is calculated as ~3 eV [57], 4.0 eV [58], and 5.5 eV [59]. The $\varepsilon(2+/0)$ transition energy was found within 0.5 eV from the VBM in Refs. [58, 57], but high in the gap around ~$E_V + 2.7$ eV in Refs. [59, 60]. Such discrepancies arise mainly due to different methodologies to correct the LDA band gap error, but may to some extent [61] also stem from different choices of the host lattice constants, which can be taken from experiment (present work), or be determined in LDA (smaller than experiment) [58], or in the generalized gradient approximation (GGA, larger than experiment) [57]. Regarding the accuracy of our present results, which are well within the range of literature data, the good agreement of the calculated optical transition energies with experimental data within our method (Ref. [29] and present work) gives us confidence that the calculated formation energies are accurate within few tenth of an eV.

### c. Anion vacancies in ZnO vs. other II-VI

We next discuss the possibility of metastable behavior and occurrence of PPC in other II-VI compounds. For ZnO, ZnS, ZnSe, and ZnTe, Fig. 5 shows both the calculated single-particle $a_1$ energy levels and the $\varepsilon(2+/0)$ transition energies of $V_X$ (X = O, S, Se, and Te). The band edges are aligned according to the calculated [62, 63] band offsets, taking also into account the present correction $\Delta E_V$ for the VBM (see section III). According to our reaction path for the light-induced transition from α- to β-behavior (Fig. 3), two conditions have to be satisfied for the occurrence of PPC: (*i*) The neutral charge state $V_X^0$ must be the initial equilibrium stable state. This means that PPC can occur only when the Fermi level is above the ε(2+/0) transition level. Since the calculated ε(2+/0) level (blue line in Fig. 5) is located around mid-gap in all II-VI compounds, PPC can occur in *n*-type, but not in *p*-type material. (*ii*) $V_X^{2+}$ must assume the conducting β-type behavior, i.e. the unoccupied $a_1^0$ has to be above the CBM (otherwise, the photoexcited electrons relax into the DLS and the vacancy returns into the non-conducting α-type $V_X^0$ state without activation). Comparing the single-particle energies of the unoccupied $a_1^0$ state of $V_X^{2+}$ (red line in Fig. 5), we see that the β-type "DLS-above-PHS" scenario is realized only in ZnO. This reflects the unusually low conduction band (high electron affinity) of ZnO. Thus, persistent photoconductivity is expected only in ZnO, provided that the Fermi level lies in the upper part of the gap above the ε(2+/0) level, e.g. in *n*-ZnO.

It is interesting to compare the chemical trends in the anion vacancy electronic states along the ZnO → ZnS → ZnSe → ZnTe series while taking into account the valence band offset [62, 63]. For $V_X^0$, we find that the occupied $a_1^2$ state comes closer to the VBM along this sequence, but shifts effectively to higher energies from ZnO to ZnTe (Fig. 5). Had the Zn dangling bonds been constructed only from Zn atomic orbitals, one would expect an alignment of the Zn-Zn metal bond on an absolute scale. Analysis of the wavefunctions shows, however, that the dangling bonds are constructed from Zn-*s*/*p* and from anion-*p* orbitals of those anions that are also neighbors to the nearest neighbor Zn atoms. Because the atomic anion-*p* orbital energy increases from O to Te, this admixture of O-*p*, S-*p*, Se-*p*, or Te-*p* contributions into the Zn-centered dangling bond causes an increase of the DLS energy on an absolute scale.

While in ZnO, we find that $V_O^0$ has locally a near-tetrahedral ($T_d$) symmetry, with one Zn atom relaxing *inwards* by a slightly larger amount than the other three equivalent Zn neighbors, in the other II-VI compounds with larger size anions, we find an unexpected symmetry-lowering relaxation mode with orthorhombic $C_{2v}$ symmetry [29], where *two* cation neighbors relax *inwards* and *two* relax *outwards*. The calculated metal-metal distances around the anion vacancy are given in Table I for the different symmetries. The total-energy difference $E[C_{2v}] - E[T_d]$ between the fully relaxed and the constrained $T_d$ symmetric structure is calculated to be −0.11, −0.32, and −0.57 eV for ZnS, ZnSe, and ZnTe, respectively [64], showing that the lower $C_{2v}$ is stabilized by a larger anion. This can be understood as a competition between the energy *gain* resulting from the lowering of the DLS level following by the inward relaxation, and the energy *loss* due to elastic deformation of *N* bonds. We see in Table I that in case of ZnO, already the near-$T_d$ configuration with *N* = 4 leads to a rather close Zn-Zn distance of 3.0 Å, while in the larger-anion compound ZnTe, the inward relaxation is limited by the elastic energy loss for *N* = 4, leading to a rather large $d_{Zn-Zn}$ = 3.89Å. In the latter situation, energy can be further lowered by constructing the DLS from only *N* = 2 dangling bond orbitals, leading to the $C_{2v}$ symmetry and a very close Zn-Zn distance of 2.48 Å. In the ionized 2+ state, no energy can be gained from a lowering of the DLS energy (DLS is unoccupied), and the Zn neighbors relax outward to a threefold coordinated, near-planar configuration (Table I).

## V. SELENIUM VACANCIES IN Cu-III-Se$_2$ CHALCOPYRITES

Figure 2b shows the angular-momentum decomposed vacancy-site local DOS of the relaxed neutral ($V_{Se}^0$) and charged ($V_{Se}^{2+}$) Se vacancy in CuInSe$_2$. We see that the relaxed $V_{Se}^0$ has an occupied *a* symmetric (*s*-like with respect to the $V_{Se}$ site) state about 2 eV *below* the VBM (Fig. 2b, middle panel). Thus, $V_{Se}^0$ assumes the α-type "DLS-below-CBM" behavior. Note that the occupied $a^2$ level of $V_{Se}^0$ in CuInSe$_2$ and CuGaSe$_2$ lies much lower in energy than the corresponding $a_1^2$ state of $V_{Se}^0$ in ZnSe, as shown in Fig. 5. The reason for this is the orbital structure of the relaxed nearest-neighbor cations: Whereas in II-VI each anion has four identical cation neighbors, in chalcopyrite, e.g. CuInSe$_2$, the anion is coordinated by two Cu atoms and two In atoms The local structure of $V_{Se}^0$ in chalcopyrite has two cations moving inwards and two moving outwards. The total-energy calculations show that the lowest energy configuration for the anion vacancy in chalcopyrite is always assumed when the dimer is formed by the group-III atoms (Ga, In) while the Cu atoms relax outward (Table I). As we found in the II-VI compounds, the wavefunction of the occupied DLS of $V_{Se}^0$ is made of *s*- and *p*-contributions at the cation site (Ga, In). Since the energies of the atomic cation *s*- and *p*-orbitals become deeper along the sequence Cu → Zn → Ga/In, this explains the very deep energies of the anion vacancy DLS in chalcopyrite (Ga-Ga dimers with deep energies) relative to II-VI compounds (Zn-Zn dimers with shallower energies). We also see why Ga-Ga and In-In dimer formation is favorable over Ga-Cu, In-Cu, or Cu-Cu combinations.

When forming the ionized 2+ state $V_{Se}^{2+}$, in which the DLS is unoccupied, both the neighboring Cu and group-III atoms relax *outwards* (Table I). As a result of the outward relaxation, the DLS of $V_{Se}^{2+}$, i.e. the $a^0$ single-particle level (cp. Figs. 2 and 5), is shifted to higher energies becoming an unoccupied resonance in the conduction band, similar to the situation in ZnO. Thus, we observe a change from α- ($V_{Se}^0$) to β- ($V_{Se}^{2+}$) behavior like in ZnO, but with the important distinction that the DLS of the neutral vacancy in chalcopyrite is located not only below the CBM, but even *below* the VBM. These results are summarized in Fig. 6a. In the $V_{Se}^+$ state where the DLS is singly occupied ($a^1$), the inward relaxation of the group-III atoms leads to the same structure as for $V_{Se}^0$ with the DLS deep below the VBM. This is seen in Fig. 6 showing schematically the calculated energy levels for $V_{Se}$ in CuInSe$_2$ (Fig. 6a) and the corresponding configuration coordinate diagram (Fig. 6b). During relaxation, the $a^1$ state of $V_{Se}^+$ becomes doubly occupied and releases one hole to the VBM, i.e. $V_{Se}^+$ is unstable against $V_{Se}^+ \rightarrow V_{Se}^0 + h$. A similar behavior of $V_{Se}^+$ was observed in CuGaSe$_2$ [29].

Figure 2b shows that $V_{Se}^0$ introduces also an empty gap state ($b^0$). The *p*-like character in the angular momentum decomposition with respect to the vacancy site (Fig. 2b, middle panel) indicates that this level originates from an anti-bonding combination of the In-centered dangling bond orbitals. When this state is occupied, $V_{Se}$ assumes the *negatively* charged states $V_{Se}^-$ and $V_{Se}^{2-}$. While the singly charged $V_{Se}^-$ has a ε(0/−) transition level above the CBM in CuInSe$_2$ and is, therefore, not stable, the doubly charged $V_{Se}^{2-}$ gives rise to a very deep *acceptor* state of $V_{Se}$ which is actually very close to the CBM, i.e. ε(0/2−) = $E_V$ + 0.97 eV = $E_C$ − 0.07 eV. In CuGaSe$_2$, which has a larger band gap ($E_g$ = 1.7eV), the deep acceptor level is located around mid-gap, i.e. $\varepsilon(0/2-) = E_V + 1.10$ eV.

The formation energies of the stable charge states of $V_{Se}$ in CuInSe$_2$, i.e. $V_{Se}^{2+}$, $V_{Se}^{0}$, and $V_{Se}^{2-}$, are shown in Fig. 4b as a function of $\Delta E_F$, indicating also the respective transition energies $\varepsilon(2+/0) = E_V + 0.02$ eV and $\varepsilon(0/2-) = E_V + 0.97$ eV. We see that only at very low values of $\Delta E_F$ (below 0.02 eV) the system is in the $V_{Se}^{2+}$ state, transforming directly into $V_{Se}^{0}$ (negative-$U$ behavior), while at $E_F$ near the CBM, $V_{Se}^{0}$ transforms directly into $V_{Se}^{2-}$ (also a negative-$U$. transition). Thus, for most of the range of $\Delta E_F$ (and actually also for realistic values of $\Delta E_F$) $V_{Se}^{0}$ is the ground state. The absence of any shallow donor states (contrasting with the expectation [65, 66] that formation of Se vacancies is responsible for the *n*-type conductivity observed after Se-poor growth [67]) and the absence of any deep transitions within the experimentally accessible range of $\Delta E_F$ indicates that the *isolated* Se vacancy would hardly affect the electronic properties of CuInSe$_2$. As discussed below, however, we find that the complex formation of $V_{Se}$ with the abundant acceptor $V_{Cu}$ plays an important role, and gives rise to gap states and to PPC.

**a. Defect physics of *p*-type PPC caused by V$_{Se}$**

The fact that the vacancy DLS can shift to energies below the VBM in Cu-III-VI$_2$ compounds leads to the possibility of persistent *hole* photoconductivity (*p*-type PPC). While we find that the PPC actually observed in experiments is caused by the vacancy complex with $V_{Cu}$, the basic physics of this metastability follows already from the behavior of the isolated $V_{Se}$, which we discuss first: As shown in Fig. 6 the reaction pathway leading to *p*-type PPC in CuInSe$_2$ starts from the ionized, β-type $V_{Se}^{2+}$ ("DLS-above-PHS") state. Note that the initial *ionized* charge state $V_{Se}^{2+}$ here is different from the initial *neutral* $V_O^0$ state required in ZnO. Illumination [19, 20, 21, 22, 23] or electron injection [68] of such CuInSe$_2$ samples create occupation of the shallow PHS (Fig. 6a, right hand side). The corresponding optical transition $V_{Se}^{2+} \rightarrow V_{Se}^{+} + h$, shown in the diagram of Fig. 6b, requires an energy $\varepsilon_o(2+/+;h)$ which practically equals the band gap energy [69]. Immediately after the optical excitation, the electron in the $V_{Se}^{+}$ state occupies the shallow PHS rather than the DLS (viz. Fig. 6), so that the β-type configuration with the large In-In distance $d_{In-In} = 5.45$ Å is stabilized. However, the DLS can capture the electron from the PHS with a small activation energy $\Delta E_1$ (Fig. 6b), because the DLS energy shifts to energies below the CBM for In-In distances smaller than ~4.8 Å. After this thermally activated transition from situation β to situation α ("DLS-below-CBM"), atomic relaxation leads, without further activation, to the formation of the In-In dimer ($d_{In-In} = 3.0$ Å, cp. Fig. 6b, left hand side). During this relaxation, which is driven by the occupation of the DLS, the DLS shifts to energies below the VBM (for $d_{In-In} < 3.9$ Å), and becomes doubly occupied, i.e. $V_{Se}^{0}$ is formed. Thus, the optical excitation initiates the overall reaction

$$V_{Se}^{2+} \rightarrow V_{Se}^{0} + 2h \tag{6}$$

after which two electrons are trapped in the deep DLS (Fig. 6a, left hand side), and two free holes are released to the valence band, leading to *p*-type PPC. The energy of the metastable ($V_{Se}^{0} + 2h$) state is shown in Fig. 4b as dashed line. Note that, in contrast to the case in ZnO, the released carriers are not bound by $V_{Se}^0$ in the metastable configuration, and no new transition energies emerge as they do in ZnO (open circles in Fig. 4a). However, the initial presence of $V_{Se}^{2+}$ requires the presence of acceptors that compensate the $V_{Se}$ double donor, and the released holes can be bound by these acceptors.

Comparing CuInSe$_2$ and CuGaSe$_2$, we see in Fig. 5 that the energy of the unoccupied $a^0$ level of $V_{Se}^{2+}$ is much closer to the conduction band edge in CuGaSe$_2$. Thus, we expect that electrons brought into the PHS after illumination or electron injection can be captured by the DLS practically without activation in CuGaSe$_2$ ($\Delta E_1 \approx 0$). The presence of an barrier $\Delta E_1$ in CuInSe$_2$ implies that the $V_{Se}^{2+}$ state could be stable against capture of electrons at low temperature. Consequently, it might be possible to freeze-out the light-induced effect in CuInSe$_2$, but not in CuGaSe$_2$. However, once $V_{Se}^{2+}$ has captured an electron, the transition into the trapped ($V_{Se}^0 + 2h$) state and the occurrence of *p*-type PPC should be very similar in CuInSe$_2$ and CuGaSe$_2$. In either material, the PPC is lost again when the metastable ($V_{Se}^0 + 2h$) state is thermally activated across the barrier $\Delta E_2$ (Fig. 6b) towards the $V_{Se}^{2+}$ state. Note that this process requires simultaneous capture of two holes from the valence band, so that it can be quite slow despite the moderate calculated barrier height $\Delta E_2 = 0.35$Å for CuInSe$_2$.

**b. Formation of ($V_{Se}$-$V_{Cu}$) vacancy complexes**

Our model for the metastability of $V_{Se}$ can account for the experimentally observed *p*-type PPC in Cu(In,Ga)Se$_2$. However, it requires that $V_{Se}^{2+}$ be the equilibrium ground state, implying that the Fermi level has to be unrealistically close to the VBM, i.e. below the $\varepsilon(2+/0) = E_V + 0.02$ eV transition level. Also, detailed studies [22, 68] revealed that for every two electrons captured by the metastable defect, one hole trap is created. (In different words, one deep acceptor state is created for every two holes released to the valence band.) Such behavior can not be explained by the *isolated* Se vacancy, but by a vacancy complex formed with $V_{Se}$ and $V_{Cu}$, the latter being an abundant acceptor in CuInSe$_2$ and CuGaSe$_2$ due to its rather low formation energy [70, 71, 72]. We find a quite large binding energy ~0.8 eV of the vacancy complex in *p*-type CuInSe$_2$ indicating that such complex formation plays an important role. According to the above discussion, the isolated $V_{Se}$ is excited into the metastable state via the transition $V_{Se}^{2+} \rightarrow V_{Se}^0 + 2h$. The underlying mechanism for the metastability, i.e. the formation ($V_{Se}^0$) and breakup ($V_{Se}^{2+}$) of In-In bonds, is still operational in the vacancy complex, leading to *p*-type PPC by the analogous reaction

$$\left(V_{Se}\text{-}V_{Cu}\right)^+ \rightarrow \left(V_{Se}\text{-}V_{Cu}\right)^- + 2h. \tag{7}$$

In contrast to $V_{Se}^0$, the negatively charged vacancy complex on the right hand side of the reaction (7) can bind one hole with the calculated binding energy $E_b = 0.27$ eV being in excellent agreement with the experimental hole trap depth 0.26 eV [68]. This mechanism for *p*-type PPC requires that the positively charged complex [left hand side of Eq. (7)] be the equilibrium ground state. The calculated $\varepsilon(+/-) = 0.31$ eV transition of the vacancy complex is much deeper in the gap than the analogous $\varepsilon(2+/0) = 0.02$ eV level of the isolated $V_{Se}$, so that the requirement of the initial $\left(V_{Se}\text{-}V_{Cu}\right)^+$ state is fulfilled in *p*-CuInSe$_2$. Therefore, we suggest that the metastability observed in actual Cu(In,Ga)Se$_2$ experiments stems from the $\left(V_{Se}\text{-}V_{Cu}\right)$ vacancy complex rather than from the isolated Se vacancy. A more detailed account of the properties of the vacancy complex will be given elsewhere.

## VI. CONCLUSION

Using first-principles electronic structure calculations, we identified the anion vacancies in II-VI and chalcopyrite Cu-III-VI$_2$ semiconductors as a class of intrinsic defects that exhibit metastable behavior. The metastability arises from a crossover of the energy of the defect localized state with respect to the host band edges when the charge state is changed. Specifically, we predict persistent *electron* photoconductivity (*n*-type PPC) caused by V$_O$ in *n*-ZnO, and persistent *hole* photoconductivity (*p*-type PPC) caused by V$_{Se}$ in *p*-CuInSe$_2$ and *p*-CuGaSe$_2$. Thus, the Se vacancy in the chalcopyrite compounds constitutes the unusual case where a donor produces *p*-type PPC. The vacancy complex $\left(V_{Se}\text{-}V_{Cu}\right)$ produces *p*-type PPC similar to the isolated Se vacancy, but, at the same time, can account for the experimental observation of a deep hole trap that appears in the metastable state. The metastability of anion vacancies provides a model for the hitherto poorly understood mechanism that causes persistent photoconductivity in CuInSe$_2$ based photovoltaic materials.

This work was supported by DOE EERE under contract No. DEAC36-98-GO10337.

Table I. Calculated distances between the metal neighbors to the anion vacancy in II-VI and I-III-VI$_2$ semiconductors. In case of the neutral vacancy, the dimerized structure ($C_{2v}$-like) is more stable than the symmetric ($T_d$-like) structure, except for ZnO. Note that the actual point group symmetry is $C_{3v}$ in ZnO, and $C_2$ in case of the chalcopyrite compounds. For comparison, the metal-metal distances in the perfect crystal are also given.

|  |  | ZnO | ZnS | ZnSe | ZnTe | CuGaSe$_2$ | CuInSe$_2$ |
|---|---|---|---|---|---|---|---|
| $V^0_{anion}$ | $T_d$ | ~ 3.0 | 3.48 | 3.63 | 3.89 | – | – |
| $V^0_{anion}$ | $C_{2v}$ (in) | – | 2.59 | 2.52 | 2.48 | 2.83 (Ga-Ga) | 3.04 (In-In) |
|  | $C_{2v}$ (out) | – | 4.52 | 4.86 | 5.29 | 4.91 (Cu-Cu) | 5.15 (Cu-Cu) |
| $V^{2+}_{anion}$ | $T_d$ | ~ 4.0 | 4.83 | 5.06 | 5.40 | 5.29 (Ga-Ga) | 5.45 (In-In) |
|  |  |  |  |  |  | 4.96 (Cu-Cu) | 5.21 (Cu-Cu) |
| host |  | 3.23 | 3.83 | 4.01 | 4.30 | 3.93 | 4.10 |

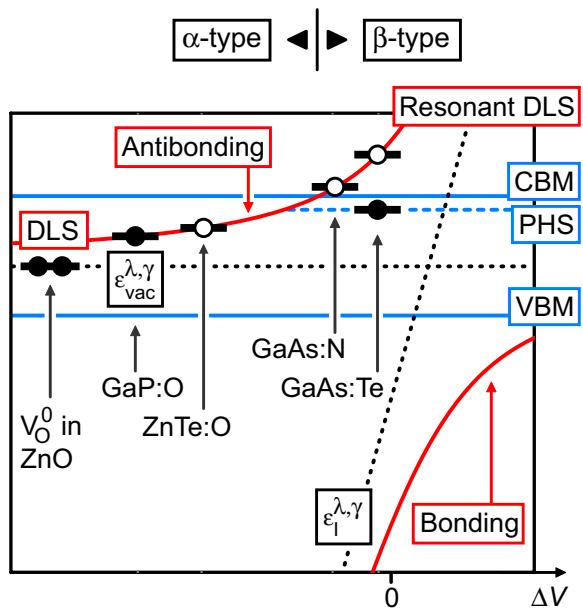

Figure 1 (color online). Schematic energy level diagram of a substitutional impurity in a semiconductor. Bonding and antibonding levels (red lines) are formed due to the interaction between the energy level $\varepsilon_{vac}^{\lambda,\gamma}$ of the pure vacancy and the atomic energy level $\varepsilon_{I}^{\lambda,\gamma}$ of the impurity atom (black dotted lines), as a function of the perturbation strength $\Delta V = \varepsilon_{I}^{\lambda,\gamma} - \varepsilon_{H}^{\lambda,\gamma}$, where $\varepsilon_{H}^{\lambda,\gamma}$ is the atomic orbital energy of the replaced host atom ($\lambda$ denotes the sublattice and $\gamma$ the symmetry representation). $\alpha$- and $\beta$-type behavior is distinguished by the relative position of the antibonding state and the host CBM (blue solid line): The defect is $\alpha$-type when the DLS is below the CBM; it is $\beta$-type when the DLS is above the CBM, and forms a PHS (blue dashed line) just below the CBM.

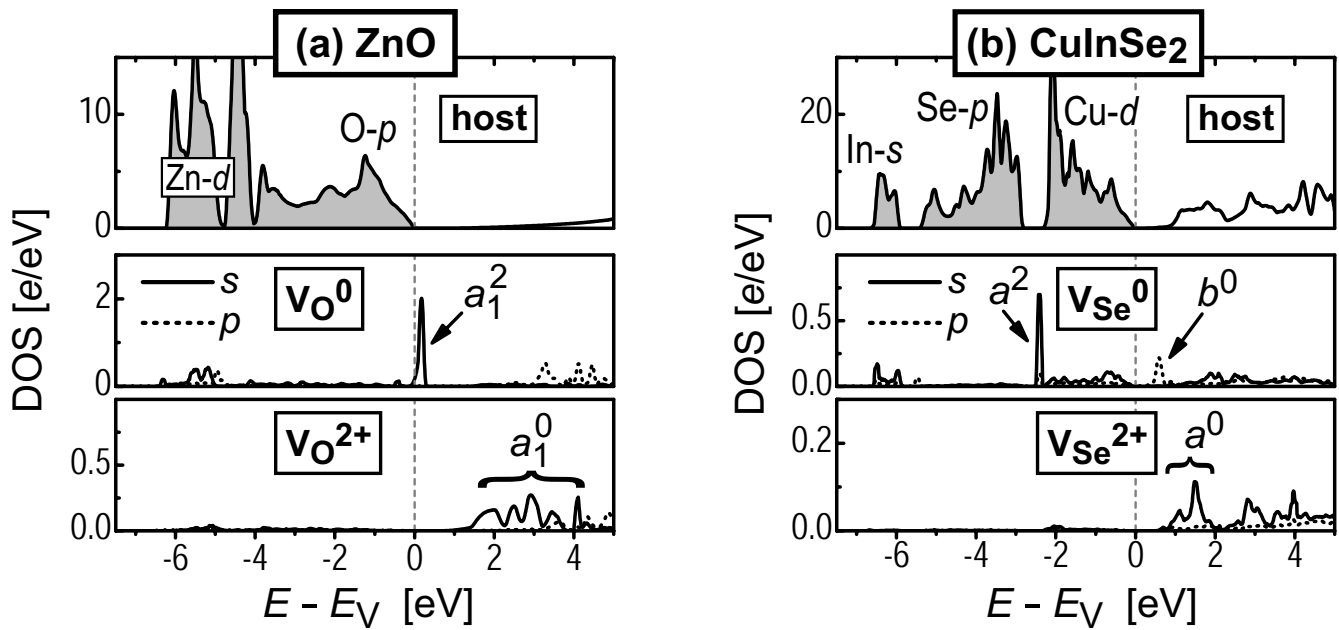

Figure 2. Calculated density of states of the host crystals and the angular momentum decomposed local DOS at the neutral ($V_X^0$) and charged ($V_X^{2+}$) vacancy site for (a) ZnO and (b) CuInSe$_2$. The occupied DLS, i.e. the $a_1^2$ ($V_O^0$) and the $a^2$ ($V_{Se}^0$) level, and the unoccupied DLS resonance in the conduction band, i.e. the $a_1^0$ ($V_O^{2+}$) and the $a^0$ ($V_{Se}^{2+}$) level, are indicated. In CuInSe$_2$, the unoccupied $b$ symmetry level of $V_{Se}^0$ is also shown.

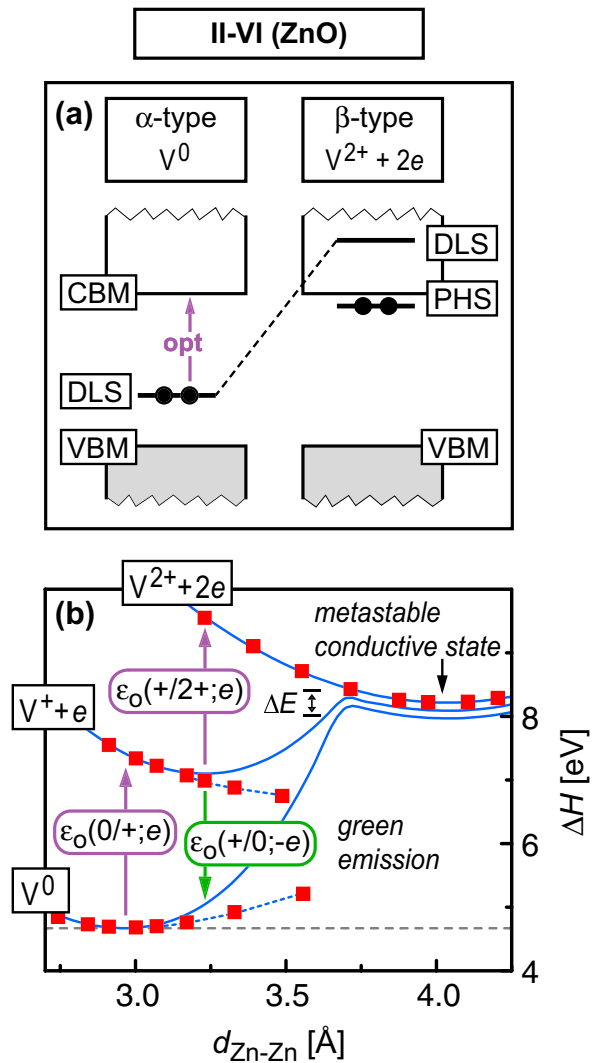

Figure 3 (color online). (a) Schematic energy diagram for $V_O$ in ZnO, causing *n*-type PPC. Electrons that occupy the DLS or the PHS are depicted as solid circles. (b) Configuration coordinate diagram, where the calculated defect formation energies $\Delta H$ (red squares) refer to oxygen-rich conditions [$\Delta\mu_O = 0$]. Note that the calculated formation energies $\Delta H[V_O^0]$ and $\Delta H[V_O^+]$ are underestimated when $d_{Zn-Zn}$ exceeds 3.2 Å (dotted lines, cp. footnote [45]). In these cases, the solid lines show a qualitative continuation of the $\Delta H$. In the β-type metastable state, two electrons are released to the shallow perturbed host state, leading to *n*-type PPC.

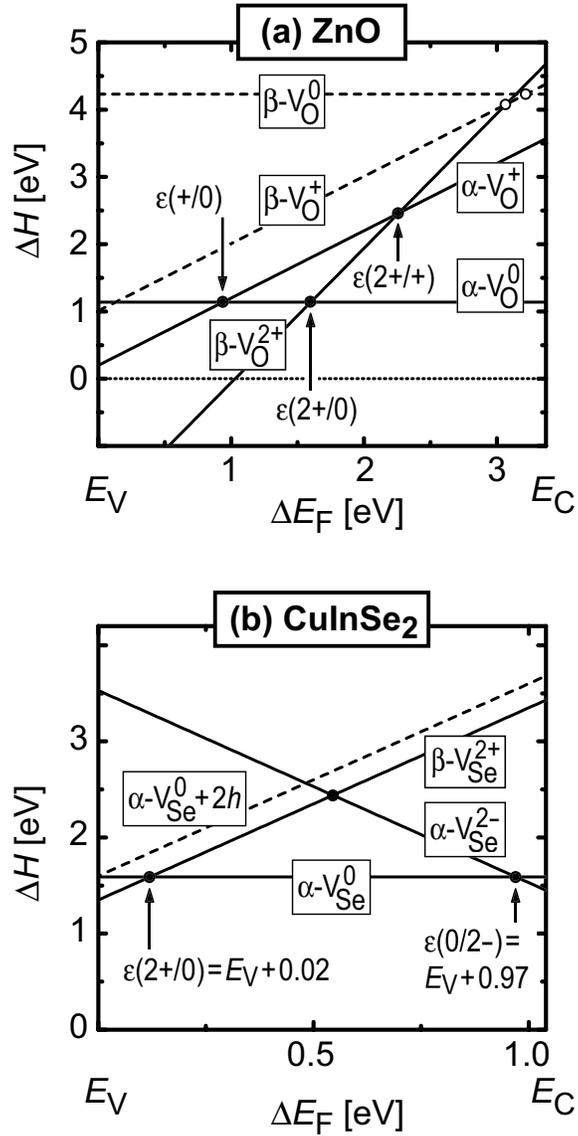

Figure 4. Formation energies $\Delta H$ of the different charge states of $V_{anion}$ in (a) ZnO and (b) CuInSe$_2$ in the limit of anion-poor conditions ($\Delta\mu_O = -3.53$ eV, $\Delta\mu_{Se} = -0.83$ eV), as a function of the Fermi level $\Delta E_F$ within the band gap. Solid lines and closed circles show the formation and transition energies of the equilibrium stable states, respectively. Note that the line for $\Delta H(V_{Se}^{2+})$ is vertically displaced by $-0.2$ eV for enhanced graphical clarity. Dashed lines and open circles show the formation and transition energies in the light induced metastable configuration. $\alpha$- and $\beta$-type behavior of the respective states is indicated.

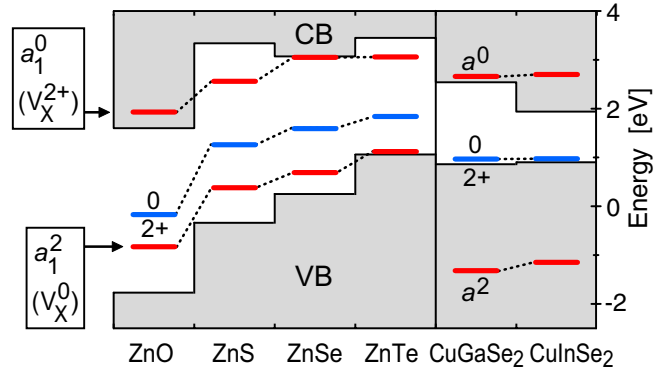

Figure 5 (color online). Single particle energies of the $a_1$ symmetric vacancy defect levels (red) and the thermal $\varepsilon(2+/0)$ transition energy (blue) for the anion vacancy in II-VI and chalcopyrite semiconductors, relative to the host band edges. The valence band offsets are taken from Refs. [62, 63]; the data shown takes into account the correction $\Delta E_V$ (cp. section III) for the position of the VBM.

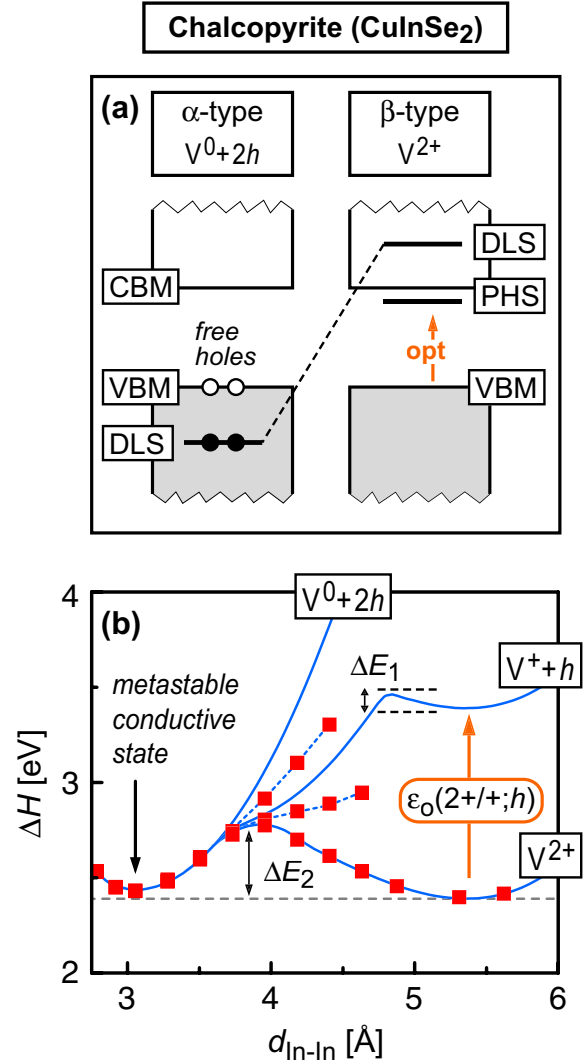

Figure 6 (color online). (a) Schematic energy diagram for $V_{Se}$ in $CuInSe_2$, causing *p*-type PPC. (b) Configuration coordinate diagram, where the calculated defect formation energies $\Delta H$ (squares) refer to Se-rich conditions ($\mu_{Se} = \mu_{Se}^{solid}$). Note that the LDA calculated formation energies for $V_{Se}^{0}$ and $V_{Se}^{+}$ are underestimated when $d_{In-In}$ exceeds 3.9 Å (dotted lines, cp. footnote [45]). In these cases, the solid lines show a qualitative continuation of the $\Delta H$. In the α-type trapped state, two free holes are released to the valence band, leading to *p*-type PPC.